# Non-equilibrium microtubule fluctuations in a model cytoskeleton

Clifford P. Brangwynne[1*†], Gijsje H. Koenderink[1**†], Frederick C. MacKintosh[2], and David A. Weitz[1,3‡]

1 School of Engineering and Applied Sciences, Harvard University, Cambridge, MA 02138, USA.
2 Department of Physics and Astronomy, Vrije Universiteit, 1081 HV Amsterdam, Netherlands
3 Department of Physics, Harvard University, Cambridge, MA 02138, USA
† These authors contributed equally to this work.
*Current address: MPI-PKS, Nöthnitzer Straße 38, Dresden, Germany
**Current address: FOM Institute AMOLF, 1098 SJ Amsterdam, Netherlands
‡e-mail: weitz@seas.harvard.edu

**PACS Numbers: 87.16.Ka, 87.68.+z , 87.16.-b**

**ABSTRACT**
**Biological activity gives rise to non-equilibrium fluctuations in the cytoplasm of cells; however, there are few methods to directly measure these fluctuations. Using a reconstituted actin cytoskeleton, we show that the bending dynamics of embedded microtubules can be used to probe local stress fluctuations. We add myosin motors that drive the network out of equilibrium, resulting in an increased amplitude and modified time-dependence of microtubule bending fluctuations. We show that this behavior results from step-like forces on the order of 10 pN driven by collective motor dynamics.**

All materials exhibit continual fluctuations due to thermal agitation; these are readily apparent in liquids and soft materials, including many biological materials. However, biological materials are unique in that they can also exhibit non-equilibrium, internal stress fluctuations that result from active processes. For example, within living cells, motor proteins utilize the chemical energy of hydrolysis of adenosine triphosphate (ATP) to exert forces that directly affect the motion of the cytoskeleton. Studies have shown that myosin motors can change the properties of reconstituted actin filament networks by sliding filaments past one another [1-7]; if the network is cross-linked, such forces can generate tension and dramatically increase the network's stiffness [8, 9]. Within living cells, motor activity can lead to diffusive-like motion [10], although little is known about its microscopic nature. The effects of these forces can be readily seen by fluorescently labeling microtubules, as shown for a Cos7 epithelial cell in Fig. 1(a); indeed, even though microtubules are the stiffest biopolymers in the cell [11], they nevertheless exhibit significant bending fluctuations, suggesting they are highly stressed due to non-equilibrium activity. This bending motion also suggests that microtubules themselves might be used as local, microrheological probes of fluctuating cytoskeletal forces, helping to elucidate the nature and magnitude of active stress fluctuations. However, this requires a quantitative understanding of microtubule shape fluctuations in response to non-equilibrium forces within the composite cytoskeleton.

In this Letter, we demonstrate that the bending dynamics of microtubules can be used to quantitatively probe force fluctuations in a reconstituted model cytoskeleton of actin filaments and myosin motors, revealing how actomyosin contractile activity leads to diffusive-like behavior similar to that observed in cells. Fourier analysis of microtubule bending dynamics shows that the bending fluctuations occur primarily on short length scales due to constraints from the elasticity of the surrounding actin network. The motor-driven fluctuations are much larger than the thermally driven ones seen in motor-free equilibrium networks; the motors give rise to diffusive-like motion that arises from step-like relaxation dynamics of collective myosin activity [12]. By directly measuring the amplitude of discrete fluctuation events, we probe the underlying forces, whose magnitude is on the order of 10 pN.

We reconstitute an active biomimetic cytoskeletal network by incorporating processive myosin II motor filaments into a network of filamentous (F-) actin in a buffer of physiological ionic strength. At high ionic strength, myosin II is present as single molecular motors, which are non-processive with a duty ratio of only 2% [13]. At 50 mM KCl, however, myosin II self-assembles into bipolar filaments each composed of several hundred motors, rendering them effectively processive [14]. Myosin II was purified from chicken skeletal muscle [15] and stored in non-filamentous form in a high ionic strength buffer (0.6 M KCl, 1 mM dithiothreitol (DTT), 50 mM phosphate, pH 6.3, 50% glycerol) at $-20^{o}$C. Fresh myosin stock solutions were prepared by dialysis against AB300 buffer (300 mM KCl, 4 mM $MgCl_2$, 1 mM DTT, 25 mM imidazole, pH 7.4). G-actin was purified from rabbit skeletal muscle [16] and stored at $-80^{o}$C in G-buffer (2 mM Tris-HCl, 0.2 mM ATP, 0.2 mM $CaCl_2$, 0.2 mM DTT, 0.005% $NaN_3$, pH 8.0). Fluorescently labeled microtubules [17], were stabilized by addition of 10 $\mu$M taxol (Sigma). Actin-myosin samples sparsely seeded with microtubules were prepared under buffer conditions that induce actin polymerization and formation of myosin thick filaments (25 mM imidazole, 50 mM KCl, 5 mM MgATP, 0.2 mM $CaCl_2$, 1 mM DTT, 0.1 mM taxol, pH 7.4). Samples were prepared by gentle mixing of buffers, taxol, an antioxidant mixture (glucose oxidase, catalase, glucose, 2-mercaptoethanol) to slow photobleaching [2], microtubules, myosin II, and actin. Upon addition of G-actin, to initiate network formation, the samples were quickly loaded into glass coverslip/microscope slide chambers with parafilm spacers and polymerized for 30 minutes at room temperature. The myosin II concentration was varied between 0.0238 and 0.476 µM, while the actin concentration was fixed at 23.8 µM (1.0 mg/ml). The average actin filament length was set at 1.5 µm using the capping protein gelsolin in a molar ratio of 1:555 to actin [18].

Microtubules embedded in reconstituted networks of F-actin without motors are nearly straight with very small thermal fluctuations. However, in the presence of active myosin motors, their behavior is strikingly different: microtubules undergo large but highly localized bending fluctuations. It appears that local, myosin-driven contractions of the actin network pull on the embedded microtubules, as sketched in the inset of Fig. 1(a). The resulting discrete, intermittent bends grow and relax quickly, as shown in the time sequence in Fig. 1(b). At high myosin concentrations, microtubules also occasionally bend on longer length scales, presumably reflecting collective contractility of the network on large length scales [19].

To characterize the spatial and temporal dependence of microtubule bending fluctuations, we decompose their shapes into Fourier modes [11, 17]. Here, the Fourier amplitudes, $a_q(t)$, reflect the instantaneous bending at a wavevector $q = \frac{n\pi}{L}$, where $n = 1,2,3...$ is the mode number and $L$ is the contour length of the filament. We calculate fluctuations in the Fourier amplitudes, $\Delta a_q(\Delta t) = a_q(t + \Delta t) - a_q(t)$, as a function of lag time, $\Delta t$, and determine $\langle \Delta a_q^2(\Delta t) \rangle$ where $\langle \ \rangle$ denotes an average over all initial times $t$. In the absence of motor proteins, we expect thermal bending fluctuations to cause the Fourier amplitudes to saturate at long timescales, $\langle \Delta a_q^2 \rangle_{max} = 2k_BT/\kappa q^2$, where $\kappa$ is the microtubule bending rigidity, $k_B$ is Boltzmann's constant and $T$ is the temperature [11]. Microtubules in motor-free networks indeed exhibit this $q$-dependence, as shown for the fluctuations averaged over many filaments in Fig. 2(a); the average persistence length, $l_p = \kappa/k_BT \approx 1$ mm (black line, Fig. 2(a)), is similar to values obtained in aqueous buffer [11]. For $q > 1$ $\mu$m the variance is independent of time and increases with $q$, reflecting the experimental noise floor [17]. As expected, the distribution of thermal bending fluctuations, $P(\Delta a_q(\Delta t))$, is well-fit by a Gaussian for all wavelengths and lag times; to quantify this, we calculate the kurtosis of the distribution, $\alpha = \frac{\langle \Delta a_q(\Delta t)^4 \rangle}{3\langle \Delta a_q(\Delta t)^2 \rangle^2} - 1$, which increases above zero for non-Gaussian distributions. Typically $\alpha \approx 0$ for all $n$ and $\Delta t$, reflecting the Gaussian distribution, as shown for an example filament in Fig. 2(b).

Addition of myosin motors (actin to myosin molar ratio of 100) dramatically increases the amplitude of the fluctuations compared to that observed in thermal equilibrium, particularly on short length scales ($q > 0.2$ $\mu$m$^{-1}$). After a lag time of only 2 seconds, these fluctuations are already significantly larger than thermal fluctuations, indicated by the black line in Fig. 2(c). However, at long length scales ($q \leq 0.2$ $\mu$m$^{-1}$), the mode amplitudes at a given lag time are similar to those of microtubules in thermal equilibrium. For higher myosin concentrations, the crossover between equilibrium and non-equilibrium behavior shifts to smaller $q$ reflecting large-scale collective contractility.

Unlike thermal fluctuations, the kurtosis of motor-driven filaments exhibits distinct non-Gaussian signatures ($\alpha \gg 0$), particularly on short length scales ($q$~1 $\mu$m$^{-1}$), as shown for the example filament in Fig. 2(d).

The time-dependence of the amplitude fluctuations of the Fourier modes also shows clear non-equilibrium signatures. We illustrate this by scaling the time evolution of the different modes onto a single master curve. In the absence of motors, we divide the data by the known saturating amplitude, $\langle \Delta a_q^2 \rangle_{max}$, and then scale them together by the apparent relaxation time of each mode, $\tau_q$. For short times, $\Delta t$, the viscoelasticity of the actin network causes each bending mode to relax sub-diffusively, $\langle \Delta a_q^2 \rangle \sim \Delta t^{0.6}$. By contrast, in the presence of myosin motors, the behavior is completely different: The modes transition to a roughly linear, or diffusive-like, time-dependence, $\langle \Delta a_q^2 \rangle \sim \Delta t$, as shown in Fig. 3. Such diffusive motion is typically associated with thermal fluctuations, but here the large amplitude of the motion clearly indicates a non-thermal origin.

The origin of the non-equilibrium behavior is the presence of the myosin motors; however, these do not act on the microtubules themselves, but on the surrounding actin network. Fluctuations in the microtubules must arise from forces exerted on them by the surrounding network. However, these forces are highly localized; evidence for this comes from our observations that the bending fluctuations of different microtubules in close proximity are not correlated. Moreover, these localized fluctuations are large, as can be seen in Fig. 1(b). We model this behavior as a point-like force, acting transversely on a rod embedded within an elastic continuum [20]. The length scale for the bend is $\ell \cong (\kappa/G)^{1/4}$, reflecting the competition between the bending energy of the microtubule, which is proportional to $\kappa$, and the elastic energy of deforming the actin network, which is proportional to the shear modulus, $G$. The amplitude of a localized bend $y(x)$, as a function of position $x$ along the filament, is $y(x) = y_0 \cdot u(x)$, where $u(x) = \left[\sin(|x|/\ell) + \cos(|x|/\ell)\right]e^{-|x|/\ell}$. The peak height is $y_0 = f\ell^3/8\kappa$, where $f$ is the applied force. The microtubule shapes are well-described by this functional form. Typical data are shown in Fig 4(a); these are fits at three different time points, to the microtubule bend shown in Fig. 1(b). From many such fits, we obtain $\ell \approx$ 1-2 $\mu$m, consistent with the peak in the amplitude of fluctuations of Fourier modes at $q \approx 0.5$ $\mu$m$^{-1}$ [Fig. 2(c)]. Using the bending rigidity obtained from Fourier analysis, $\kappa \approx 4 \times 10^{-24}$ Nm$^2$, this yields $G \approx$ 1 Pa, in agreement with macroscopic rheological measurements (not shown). From the maximum bending amplitudes, $y_0^{max}$, we obtain a distribution of maximum forces, $f^{max}$, with an average around 10 pN and a tail extending up to 40 pN, as shown in Fig. 4(b) (N=20-50 events). This force scale suggests that several myosin motors produce the bend, consistent with the known duty ratio of myosin at mM ATP concentrations [13]. The distribution of forces is independent of the myosin concentration.

By fitting these localized bends, we track the temporal evolution of $y_0(t)$. The amplitude always shows a growth phase followed by a decay phase, occasionally separated by a phase where it remains almost constant. The growth phase is frequently exponential, with an average growth velocity of 1 μm/sec, as illustrated by the examples in Fig. 4(c). The relaxation phase is also approximately exponential with a decay time of 0.01-2 sec. Such an exponential decay would occur if the force was abruptly released, and the relaxation is a material property of the F-actin network; then the relaxation time is $\tau \cong \eta \ell / k_{eff}$, where $\eta \cong 1$ Pa*s, and $k_{eff} \cong 10$ pN/μm. This gives $\tau \approx 0.1\text{-}1$ sec, in accord with our observations. Consistent with this, we observe no systematic dependence of the growth and decay times on bending amplitude or on myosin concentration (not shown). These rapid relaxations, and corresponding instantaneous motor release are also consistent with recent observations of particle motion in active gels [9].

This rapid, step-like behavior of myosin-induced force fluctuations provides an explanation of the microscopic origin of the diffusive dynamics of the Fourier modes: The myosin-driven force turns on at a time $t$, leading to a rapid displacement of the microtubule; then, the motors stall and the force remains approximately constant for a variable period $T$; finally, the motors release and the force goes rapidly to zero. On time scales longer than this release time, the microtubule shape is approximately $y(x,t) \cong u(x) f(t) / k_{eff}$. For time scales short compared with $T$, or frequencies greater than $1/T$, this results in a frequency-dependent spectrum of modes $\langle |a_{q,\omega}|^2 \rangle \propto \frac{1}{\omega^2}$, with $\langle \Delta a_q^2(\Delta t) \rangle \sim \Delta t$ [12]. Thus, the diffusive but non-thermal cytoskeletal fluctuations can be understood in terms of the properties of collective myosin motor assemblies.

These results show that microtubules are valuable local probes of cytoskeletal force fluctuations, and can be used to elucidate the origin and microscopic nature of these fluctuations. The microtubule bending fluctuations reveal that processive motors, usually associated with sustained and directed force generation, can also lead to fluctuating dynamics, due to their collective properties. In a network with a viscoelastic response, these fluctuations result in motion that is remarkably diffusive-like, but is nevertheless distinctly non-thermal. Although here we focus on a well-defined *in vitro* model cytoskeleton, similar microtubule bending dynamics are also observed *in vivo* in living cells [21]; this suggests that collective motor activity also dominates intracellular fluctuations. This motion is not restricted to microtubules; indeed, similar diffusive-like fluctuations have been reported for embedded tracer particles [10]; this makes any measure of microrheological properties in a cell that assumes purely equilibrium fluctuations highly suspect. This must be incorporated in our understanding of all apparently thermal-like fluctuations in living cells.

We thank T. Mitchison and Z. Perlman for their kind donation of tubulin and assistance with fluorescent labeling, Z. Dogic for myosin II and his kind hospitality, and D. Mizuno, C.F. Schmidt, and L. Mahadevan for helpful discussions. This work was supported by the NSF (DMR-0602684 and CTS-0505929), the Harvard MRSEC (DMR-0213805), the Harvard IGERT on Biomechanics (DGE-0221682), and FOM/NWO. G.H.K. is supported by a European Marie Curie Fellowship (FP6-2002-Mobility-6B, Contract No. 8526). C.P.B. acknowledges the hospitality of the Vrije Universiteit.

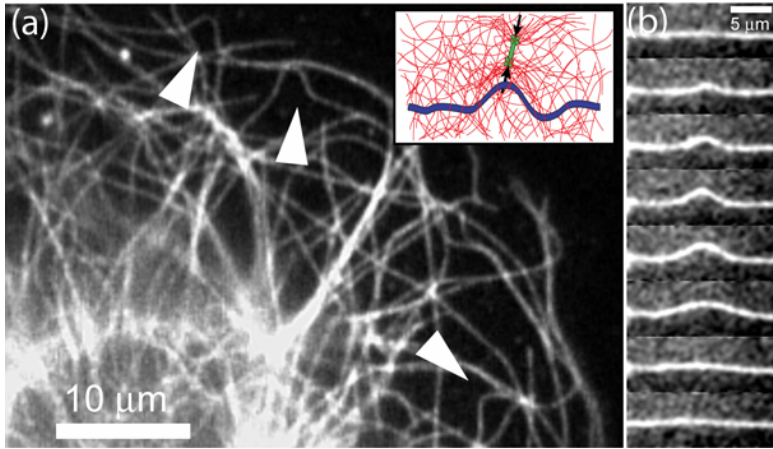

FIG. 1 (color online). (a) Fluorescently-labeled intracellular microtubules showing highly bent shapes. The inset is a schematic showing a microtubule bending under the action of myosin motors pulling on the actin network. (b) Localized microtubule bending fluctuations in the motor-driven *in vitro* network, time between frames is 78 msec.

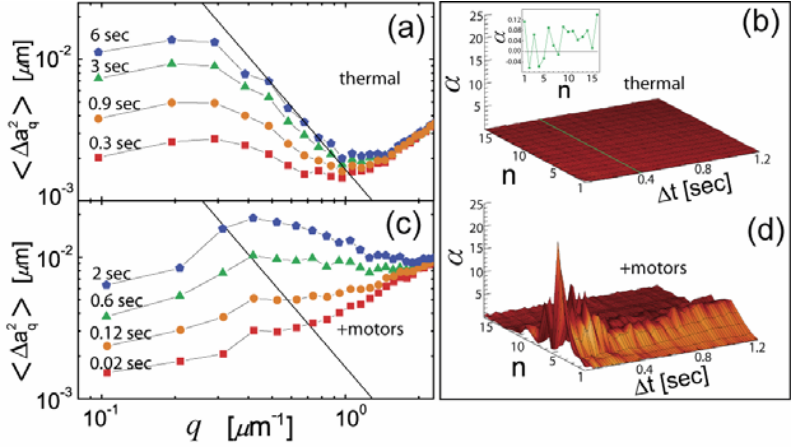

FIG. 2 (color online). (a) The saturating variance of thermal bending fluctuations scales as $1/q^2$, with $l_p \sim 1$ mm (solid line). (b) The non-Gaussian parameter, $\alpha$, for thermal bending fluctuations, as a function of lag time and mode number. Upper inset shows a higher magnification cut along $\Delta t = 0.4$ sec. (c) Motor proteins (1:100) lead to enhanced bending fluctuations on short length scales compared to the saturating thermal values. (d) $\alpha$, for motor-driven fluctuations. Example filaments in (b) and (d) have $L \sim 30\,\mu$m. Data in (a) and (c) are averages over 10-20 filaments.

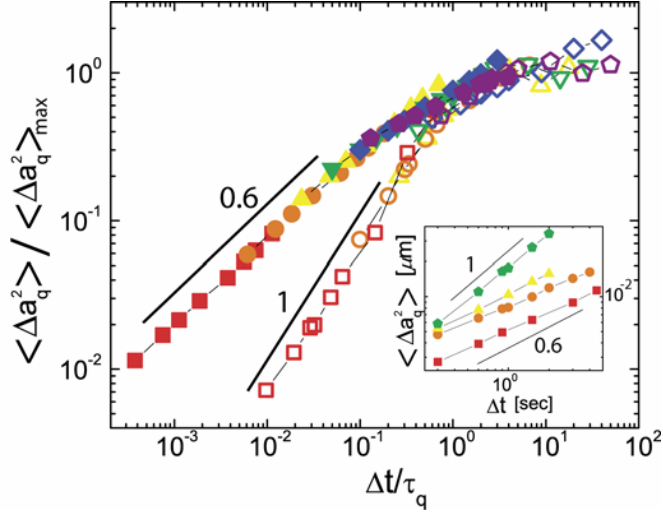

FIG. 3 (color online). Sub-diffusive dynamics, $\sim \Delta t^{0.6}$, for thermally fluctuating microtubules, solid symbols. Squares, circles, triangles, inverted triangles, diamonds, pentagons, correspond to $q = 0.097, 0.194, 0.291, 0.388, 0.484, 0.58$ $\mu$m$^{-1}$, respectively. At high myosin concentration, the fluctuations evolve diffusively with lag time, $\sim \Delta t$, open symbols. Squares, circles, triangles, inverted triangles, diamonds, pentagons, correspond to $q = 0.18498, 0.36996, 0.55494, 0.73992, 0.92491, 1.10989$ $\mu$m$^{-1}$, respectively. Lower inset shows the transition to diffusive time-dependence with increasing myosin concentration; $q \sim 0.29$ $\mu$m$^{-1}$, squares, circles, triangles, pentagons correspond to no myosin, 1:200, 1:100, and 1:50 (molar ratio, myosin:actin).

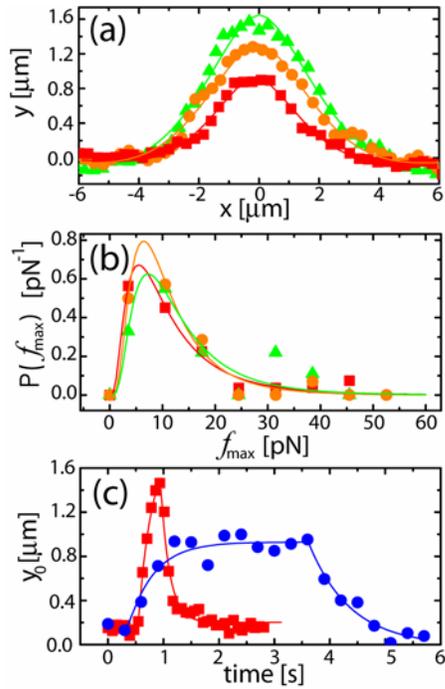

FIG. 4 (color online). (a) Shape of a section of the microtubule shown in Fig. 1(b), along with fits to the predicted shape. (b) Distribution of maximum forces obtained for various myosin concentrations (squares: 1:50; circles: 1:100; triangles: 1:200). (c) Examples of the time-dependence of the bending amplitude.

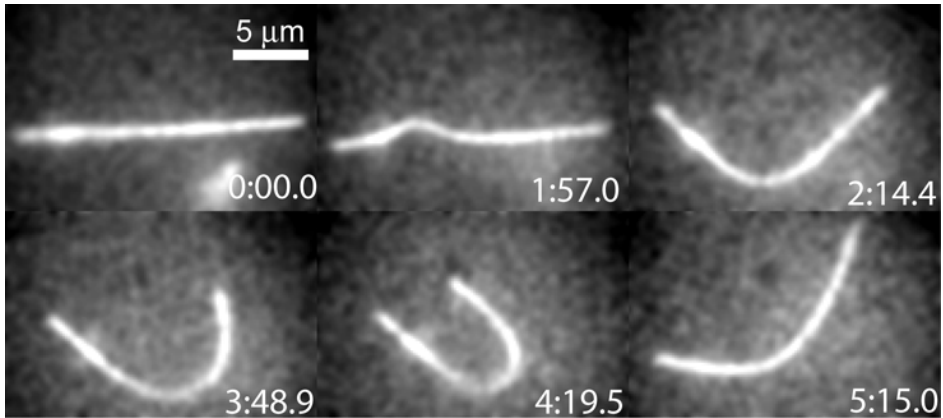

EPAPS Supplementary Fig.
Microtubules in reconstituted actin-myosin networks with a high myosin concentration (1:50) also undergo bending fluctuations on longer length scales, likely due to collective contractility of the network arising from transient motor cross-linking.